\documentclass[twocolumn,amsmath,amssymb,superscriptaddress,prb]{revtex4}

\usepackage{subscript}
\usepackage{graphicx}
\usepackage{multirow}
\usepackage[euler]{textgreek}

\usepackage[breaklinks]{hyperref}
\hypersetup{
    bookmarks=false,
    pdfstartview={FitH},
    colorlinks=true,
    linkcolor=blue,
    citecolor=blue,
    urlcolor=blue
}

\begin{document}

\title{Observation of weakened V-V dimers in the monoclinic metallic
phase of strained VO$_2$}

\author{J.~Laverock}
\affiliation{H.~H.~Wills Physics Laboratory, University of Bristol, Tyndall
Avenue, Bristol BS8 1TL, UK}
\affiliation{Department of Physics, Boston University, 590 Commonwealth Avenue,
Boston, MA 02215, USA}

\author{V.~Jovic}
\affiliation{School of Chemical Sciences and MacDiarmid Institute for Advanced
Materials and Nanotechnology, University of Auckland, Auckland 1142, New
Zealand}

\author{A.~A.~Zakharov}
\affiliation{MAX-lab, Lund University, SE-221 00 Lund, Sweden}

\author{Y.~R.~Niu}
\affiliation{MAX-lab, Lund University, SE-221 00 Lund, Sweden}

\author{S.~Kittiwatanakul}
\affiliation{Department of Materials Science and Engineering, University of
Virginia, Charlottesville, VA 22904, USA}

\author{B.~Westhenry}
\affiliation{H.~H.~Wills Physics Laboratory, University of Bristol, Tyndall
Avenue, Bristol BS8 1TL, UK}

\author{J.~W.~Lu}
\affiliation{Department of Materials Science and Engineering, University of
Virginia, Charlottesville, VA 22904, USA}

\author{S.~A.~Wolf}
\affiliation{Department of Materials Science and Engineering, University of
Virginia, Charlottesville, VA 22904, USA}
\affiliation{Department of Physics, University of Virginia, Charlottesville, VA
22904, USA}

\author{K.~E.~Smith}
\affiliation{Department of Physics, Boston University, 590 Commonwealth Avenue,
Boston, MA 02215, USA}
\affiliation{School of Chemical Sciences and MacDiarmid Institute for Advanced
Materials and Nanotechnology, University of Auckland, Auckland 1142, New
Zealand}

\begin{abstract}
Emergent order at mesoscopic length scales in condensed matter can provide
fundamental insight into the underlying competing interactions and their
relationship with the order parameter. Using
spectromicroscopy, we show that mesoscopic stripe order near the metal-insulator
transition (MIT) of strained VO$_2$ represent periodic modulations in both
crystal symmetry and V-V dimerization. Above the MIT, we unexpectedly find the
long range order of V-V dimer strength and crystal symmetry become dissociated
beyond $\approx 200$~nm,
whereas the conductivity transition proceeds homogeneously in a narrow
temperature range.
\end{abstract}

\maketitle
The recent observations of the separation of the electronic and structural
transitions of VO$_2$ have shed new light on the microscopic mechanism of the
metal-insulator transition (MIT) of this exemplar MIT
material.\cite{laverock2014,arcangeletti2007et,kim2008et} Bulk
VO$_2$ at low temperature exists in a strongly dimerized (between neighboring V
ions) and insulating monoclinic structure (the $M_1$ phase), which transitions
to the high symmetry tetragonal rutile ({\em R}) phase at 65~$^{\circ}$C
concomitant with a huge change in the conductivity of 4 orders of
magnitude.\cite{morin1959et,goodenough1973} The advantageous properties
of this transition,\cite{jeong2013,cavalleri2001,park2013} have placed VO$_2$ at
the forefront of exploitable new
technologies,\cite{jeong2013,nakano2012et} particularly when
married with recent advances in its preparation and
growth.\cite{muraoka2002,cao2009} However, the coincidence of these large
structural and electronic transitions preclude the unambiguous untangling of the
mechanism behind the transition, leading to decades of debate as to the whether
the MIT is driven by structural instabilities towards dimerization, or as a
consequence of the effects of strong electron
correlation.\cite{goodenough1973,haverkort2005,zylbersztejn1975} In this
context, the details of the low symmetry monoclinic-like metallic phase(s) that
develops intermediate between $M_1$ and {\em R} offer a unique and crucial
insight into these questions.

In addition to the well-known $M_1$ and {\em R} phases, VO$_2$ is host to a rich
set of structural phases in close proximity to its solid-state triple
point,\cite{park2013} which are accessible by alloying and/or
strain,\cite{muraoka2002,pouget1974,pouget1975et} and are
suggestive of a highly contoured energy landscape with competing structural
instabilities.\cite{park2013,gu2010,budai2014} For example, whereas each V ion
is paired via dimerization and twisted about its dimer axis in the $M_1$
structure,\cite{goodenough1973} the related $M_2$ and triclinic {\em T}
structures both contain alternating strongly and weakly dimerized V
bonds.\cite{pouget1974} The similar magnitude of the band gaps in all three of
these insulating structures, despite differing V-V bonding patterns, has been
taken as evidence that electron correlations dominate the MIT.\cite{huffman2017}
Other apparently distinct phases have also been proposed either as intermediate
in the MIT or in close proximity, such as monoclinic
metallic\cite{laverock2014,arcangeletti2007et,kumar2014} and insulating
rutile phases.\cite{kumar2014,laverock2012,kittiwatanakul2014et}
The key to
understanding the origin of the MIT in VO$_2$ may lie with a microscopic
understanding of such phases, particularly whether V-V dimers remain strongly
paired despite remaining metallic; recent measurements suggest the conductivity
is intimately linked to the ``twisting angle'' of the V-V pairs.\cite{yao2010}

\begin{figure}[t!]
\begin{center}
\includegraphics[width=1.0\linewidth,clip]{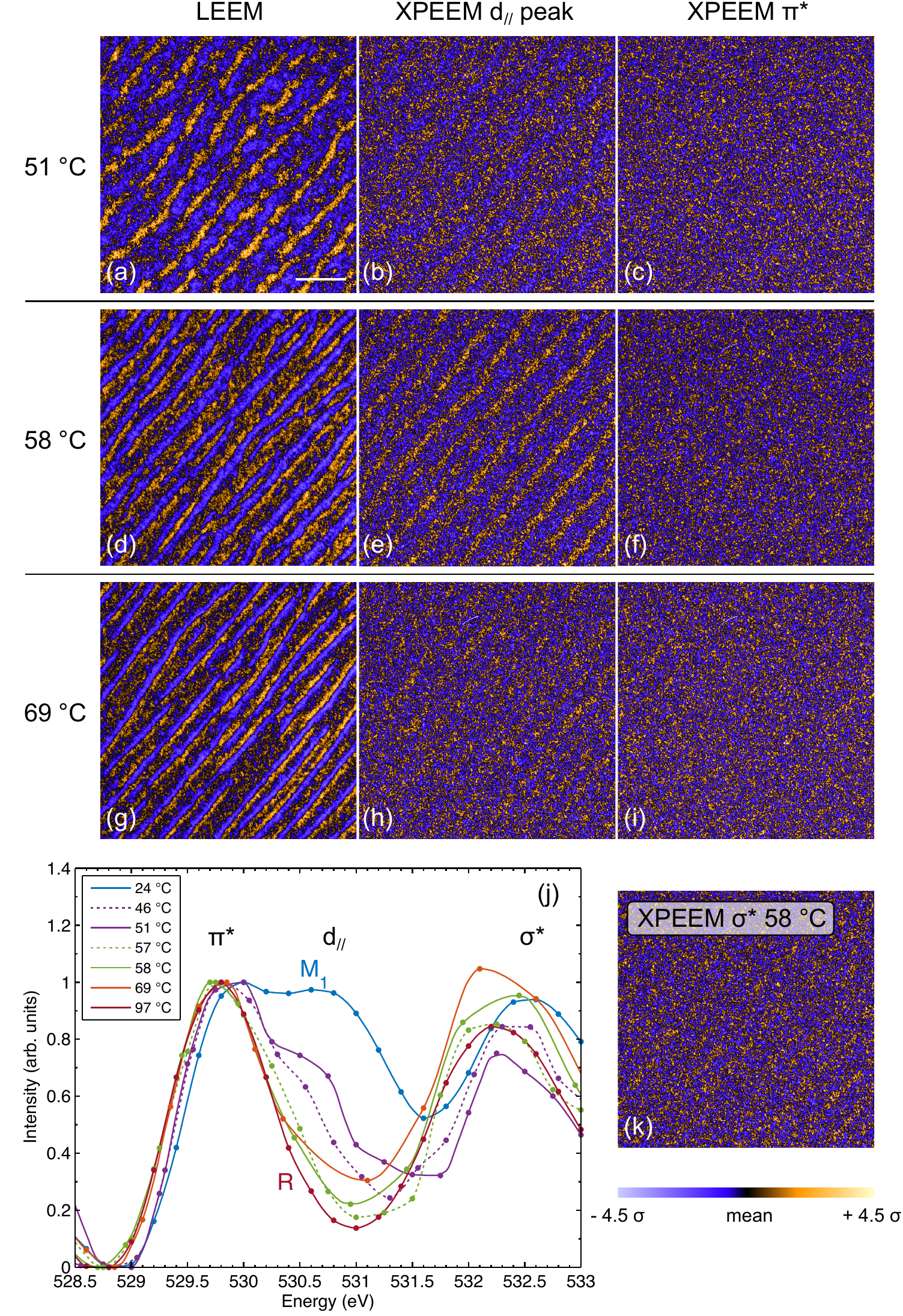}
\end{center}
\vspace*{-0.2in}
\caption{\label{f:fig1} 
LEEM and XAS-PEEM images of VO$_2$(110) at select temperatures through the MIT;
(a-c)~51~$^{\circ}$C, (d-f)~58~$^{\circ}$C, (g-i)~69~$^{\circ}$C.
At each temperature, three images are shown; LEEM,
XPEEM at the $d_{\|}$ photon energy (531~eV), and
XPEEM at the $\pi^*$ photon energy (529.7~eV). (j)~Spatially averaged
O~{\em K} edge XAS obtained by integrating the energy-dependent XAS-PEEM images
over their field of view (approx.\ 100~{\textmu}m$^2$). (k)~Example at
58~$^{\circ}$C of the recovery of XPEEM image contrast at the $\sigma^*$
absorption feature (532.5~eV). The color scale of each image is normalized to
$\pm 4.5\sigma$ from the mean LEEM or XPEEM ($d_{\|}$) values. Scale
bar 1~{\textmu}m.}
\end{figure}

Here, we employ spatially-resolved x-ray absorption spectroscopy (XAS) by
imaging the secondary electron yield using photoemission electron microscopy
(XPEEM) in concert as a spectromicroscopic technique (XAS-PEEM),
which we apply in detail
to the intermediate region of the MIT. XAS at the O~{\em K} edge in VO$_2$ is
sensitive to both V-V dimerization through the so-called $d_{\|}$ peak, which
arises due to the absorption of x-rays into unoccupied dimer ($a_{1g}$)
orbitals, and to the metallicity of VO$_2$ via the leading edge of the spectrum,
which exhibits $\sim 0.2$~eV shift between insulating and metallic
phases.\cite{abbate1991et,laverock2012b}

High quality, 300~nm thin films of VO$_2$ were grown on (110)-oriented rutile
TiO$_2$ substrates, leading to biaxial in-plane strain (tensile along $c_{\rm
R}$ and compressive along $[110]_{\rm
R}$).\cite{west2008,kittiwatanakul2013,supplementary} Spectromicroscopic
measurements were performed using low energy electron microscopy\cite{bauer1994}
(LEEM, 7~eV electron energy) and
XAS-PEEM\cite{bauer2001,supplementary}
(1.7~eV) at the SPELEEM endstation of Beamline I311,
MAX-lab (Lund, Sweden), with x-ray energy and position resolutions of 0.2~eV and
40~nm, respectively,\cite{supplementary} and diffraction
patterns were regularly monitored for electron or x-ray
damage.\cite{laverock2015} Depth sensitivities of LEEM and XAS-PEEM are 1 and
5~nm, respectively. Crucially, our biaxially strained VO$_2$(110) thin
films serve two purposes, allowing both the intermediate region of the phase
diagram to be explored, and the different coexisting phases to be probed
independently but at the same time.

Figure \ref{f:fig1} summarises the main experimental results: LEEM and XAS-PEEM
images are shown at three important stages in the MIT of VO$_2$(110)
[Fig.~\ref{f:fig1}(a-i)], alongside spatially-averaged O {\em K} edge XAS
[Fig.~\ref{f:fig1}(j)]. The three temperatures chosen
represent\cite{laverock2014} (i) 51~$^{\circ}$C: insulating, emerging
rutile-like stripes; (ii) 58~$^{\circ}$C: stripes fully established, both
metallic and insulating monoclinic-like phases coexist; and (iii)
69~$^{\circ}$C: fully metallic, small stripes of monoclinic-like symmetry
persist.

Below the MIT at 24~$^{\circ}$C, the characteristic $M_1$ phase spectrum is
observed in Fig.~\ref{f:fig1}(j), consisting of a peak at 530.0~eV due to
$\pi^*$ states, followed by a prominent $d_{\|}$ peak at 530.7~eV due to V-V
dimerization.\cite{abbate1991et,laverock2012b}
The complete suppression of the $d_{\|}$ peak above the MIT (97~$^{\circ}$)
is due to the absence of V-V dimers in the {\em R} phase,
and the shift downwards by 0.2~eV of the leading edge arises due to its
metallicity. At intermediate temperatures, some excess intensity exists at the
$d_{\|}$ peak, suggestive of the presence of V-V dimers.
Spectra recorded at temperatures $\geq 57$~$^{\circ}$C show the same leading
edge and $\pi^*$ spectra as rutile VO$_2$, indicating the sample predominantly
contains metallic VO$_2$, despite the persistence of rutile and monoclinic
stripes up to 87~$^{\circ}$C. However, at intermediate temperatures of
51~$^{\circ}$C and below, the absorption onset follows the insulating monoclinic
shape, and the $\pi^*$ feature appears intermediate between the $M_1$ and {\em
R} phase spectra. These results are in good agreement with our previous LEEM and
photoemission measurements on the same VO$_2$(110) system, where fully metallic
photoemission spectra were observed at 61~$^{\circ}$C and
above.\cite{laverock2014}

Structural stripes oriented along the $c_{\rm R}$ axis
are clearly observed in the LEEM images, Figs.~\ref{f:fig1}(a,d,g). The contrast
in the LEEM is due to diffraction contrast, in which
bright intensity corresponds to the rutile-like phase.\cite{laverock2014}
Corresponding contrast is visible in the XAS-PEEM images recorded at the
$d_{\|}$ energy shown in Fig.~\ref{f:fig1}(b,e,h), but is absent in XAS-PEEM
at the $\pi^*$ energy [Fig.~\ref{f:fig1}(c,f,i)].\cite{contrast}
This strong contrast gradually weakens before
becoming faintly visible again just above the $\sigma^*$ states [e.g.~at
58~$^{\circ}$C in Fig.~\ref{f:fig1}(k)].

\begin{figure}[t!]
\begin{center}
\includegraphics[width=0.9\linewidth,clip]{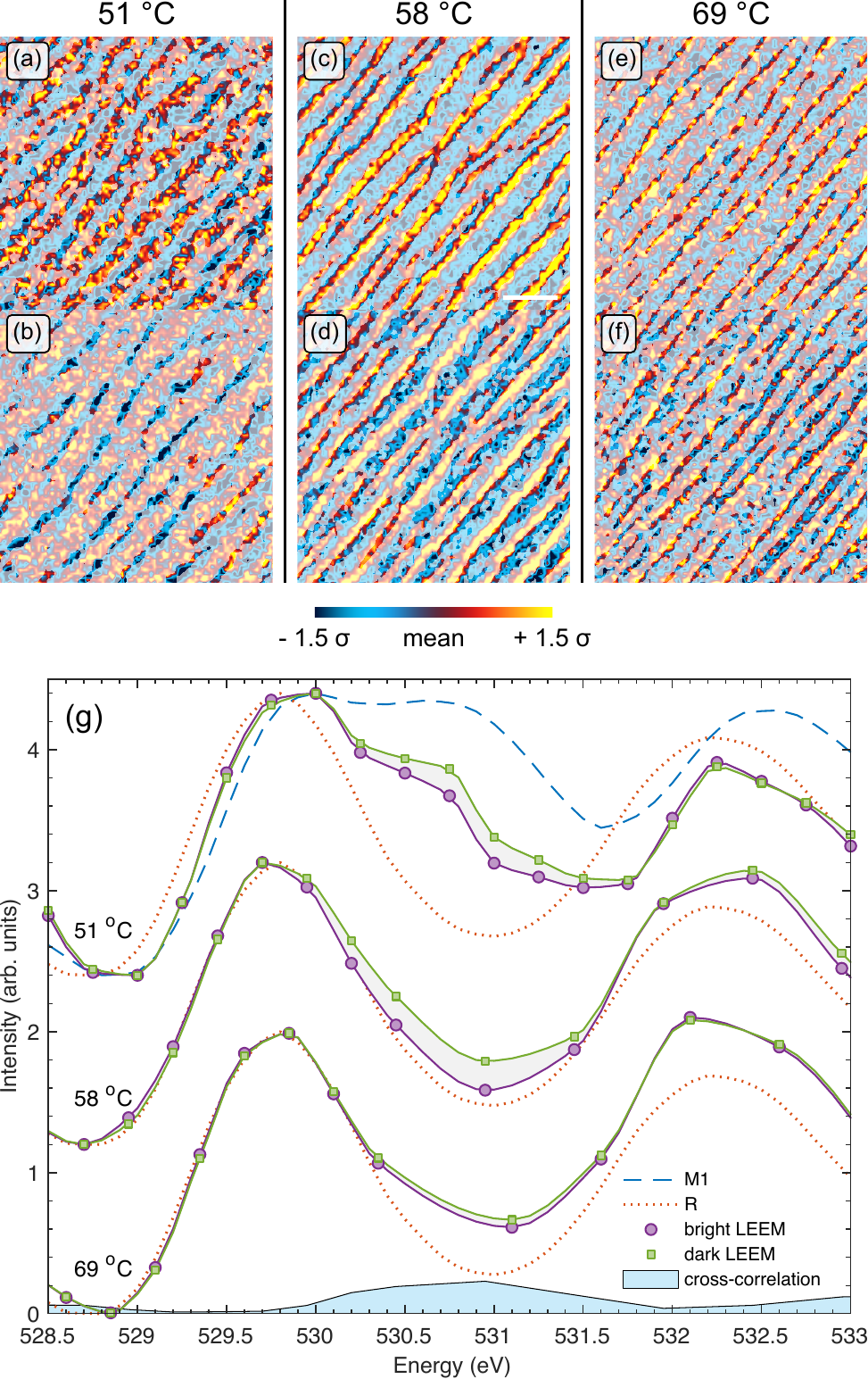}
\end{center}
\vspace*{-0.2in}
\caption{\label{f:masks} 
XAS-PEEM spectra at select temperatures obtained as a result of applying masks
to the energy-dependent XPEEM images. The dark and bright LEEM masks are shown
at each temperature in (a-f), overlaid on the corresponding $d_{\|}$ peak XPEEM
images. The thresholds used were 40\%, 30\% and 30\% for dark masks in
(a,c,e), and 15\%, 33\% and 40\% for the bright masks in
(b,d,f). Transparent regions are those
that contribute to the XAS-PEEM spectra. (g) XAS-PEEM spectra obtained by
applying the masks, vertically offset for clarity, and shown alongside reference
$M_1$ and {\em R} phase spectra. Also shown is the r.m.s.\ magnitude of the
energy-dependent cross-correlation function at 58~$^{\circ}$C. Scale bar
1~{\textmu}m.}
\end{figure}

\begin{figure}[t!]
\begin{center}
\includegraphics[width=1.0\linewidth,clip]{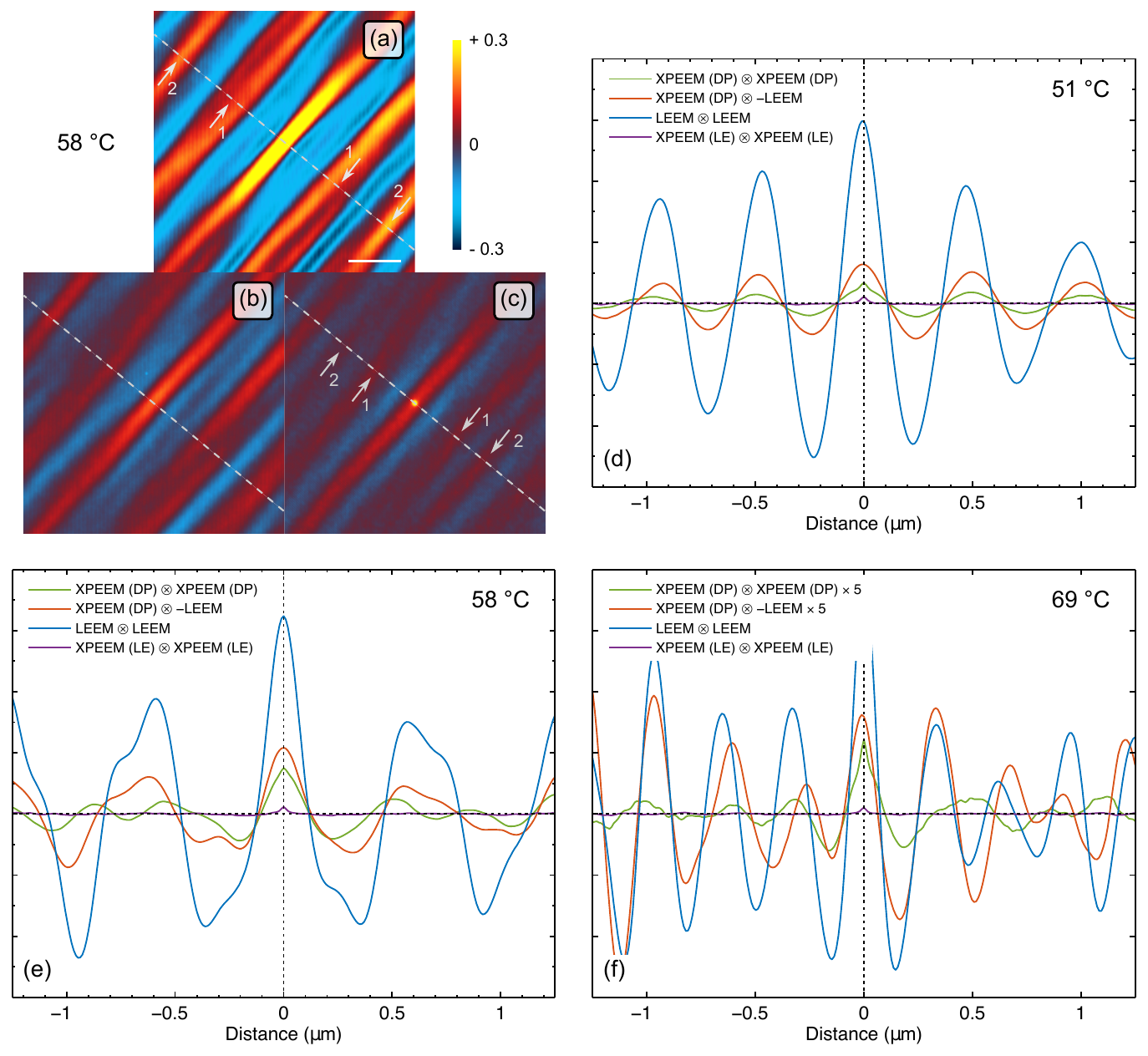}
\end{center}
\vspace*{-0.2in}
\caption{\label{f:fig3} 
Auto- and cross-correlation of XAS-PEEM and LEEM at select temperatures.
(a-c)~Examples of 2D image correlation at 58~$^{\circ}$C: (a)autocorrelation
of LEEM, (b)~cross-correlation of (inverse) LEEM and XAS-PEEM at the
$d_{\|}$ photon energy (530 -- 531 eV), and (c)~autocorrelation of XAS-PEEM at
the $d_{\|}$ photon energy. The origin is at the centre of each image.  Arrows
in (a,c) indicate the first and second satellites. Scale bar 0.5~{\textmu}m.
(d,e,f)~Auto- and cross-correlation
profiles at the three temperatures, obtained by integrating
along the stripe axis.}
\end{figure}

We now quantitatively compare XAS-PEEM and LEEM to understand the nature of the
V-V pairing in the monoclinic metallic phase. We extract the XAS spectra
corresponding to the two LEEM structures by constructing a mask from the LEEM
by thresholding its intensity.\cite{supplementary} The results are
summarised in Fig.~\ref{f:masks}. As an
example of the process at 58~$^{\circ}$C, Fig.~\ref{f:masks}(c,d) shows the
XAS-PEEM image in the $d_{\|}$ energy window, overlaid with two LEEM masks,
corresponding to the dark LEEM intensities [monoclinic-like phase,
Fig.~\ref{f:masks}(c)] and bright LEEM intensities [rutile-like phase,
Fig.~\ref{f:masks}(d)]. The agreement between the stripe structure observed in
LEEM and XAS-PEEM at the $d_{\|}$ photon energy is excellent, confirming the
contrast in both images has the same spatial origin. Similar results are found
at 51~$^{\circ}$C [Fig.~\ref{f:masks}(a,b)]
and 69~$^{\circ}$C [Fig.~\ref{f:masks}(e,f)].

The extracted XAS-PEEM spectra are shown in
Fig.~\ref{f:masks}(g), alongside the insulating $M_1$ and metallic {\em R}
phase XAS spectra. At 51~$^{\circ}$C, there is clear
contrast near the $d_{\|}$ photon energy, however the intensity is
intermediate between $M_1$ and {\em R}, suggesting both monoclinic-like and
rutile-like regions of the sample retain V-V dimer bonds to some extent.
The ``dark LEEM''
spectrum, representing monoclinic-like VO$_2$(110), was
obtained using a threshold of 40\% of the LEEM pixels at 51~$^{\circ}$C,
though this spectrum
hardly changes at a threshold of 1\%, indicating that $M_1$ VO$_2$ has already
been destroyed. Similarly, the ``bright LEEM'' spectrum does not approach the
{\em R} phase spectrum down to 1\% threshold. 
In particular, the shapes of the leading edges of the 51~$^{\circ}$C spectra
imply this is a spectrally distinct phase of VO$_2$.\cite{supplementary}
These observations demonstrate
that even at this temperature, the intermediate ``mixed'' phase of strained
VO$_2$(110) contains neither $M_1$ nor {\em R} phase VO$_2$, and that the
contrast in both LEEM and XAS-PEEM is more complex.

At 58~$^{\circ}$C, both extracted spectra show the same
leading edge and $\pi^*$ structures, supporting the homogeneity of the
metallicity. On the other hand, clear contrast between the two spectra is
observed at the $d_{\|}$ peak, supported by the peak in the r.m.s.\
cross-correlation coefficient.\cite{supplementary}
The ``bright LEEM'' spectrum approaches the
metallic rutile spectrum at all energies, whereas the ``dark LEEM'' spectrum
exhibits weak enhancement in the absorption coefficient near the $d_{\|}$
peak, and does not approach the $M_1$ phase spectrum at thresholds down to 1\%.
The monoclinic metallic phase is therefore spectrally distinct from the $M_1$
phase, consisting of weakened V-V dimers. Similar results are found at
69~$^{\circ}$C, but with reduced contrast and an overall
reduction in the intensity at the $d_{\|}$ photon energy.

\begin{figure}[t!]
\begin{center}
\includegraphics[width=1.0\linewidth,clip]{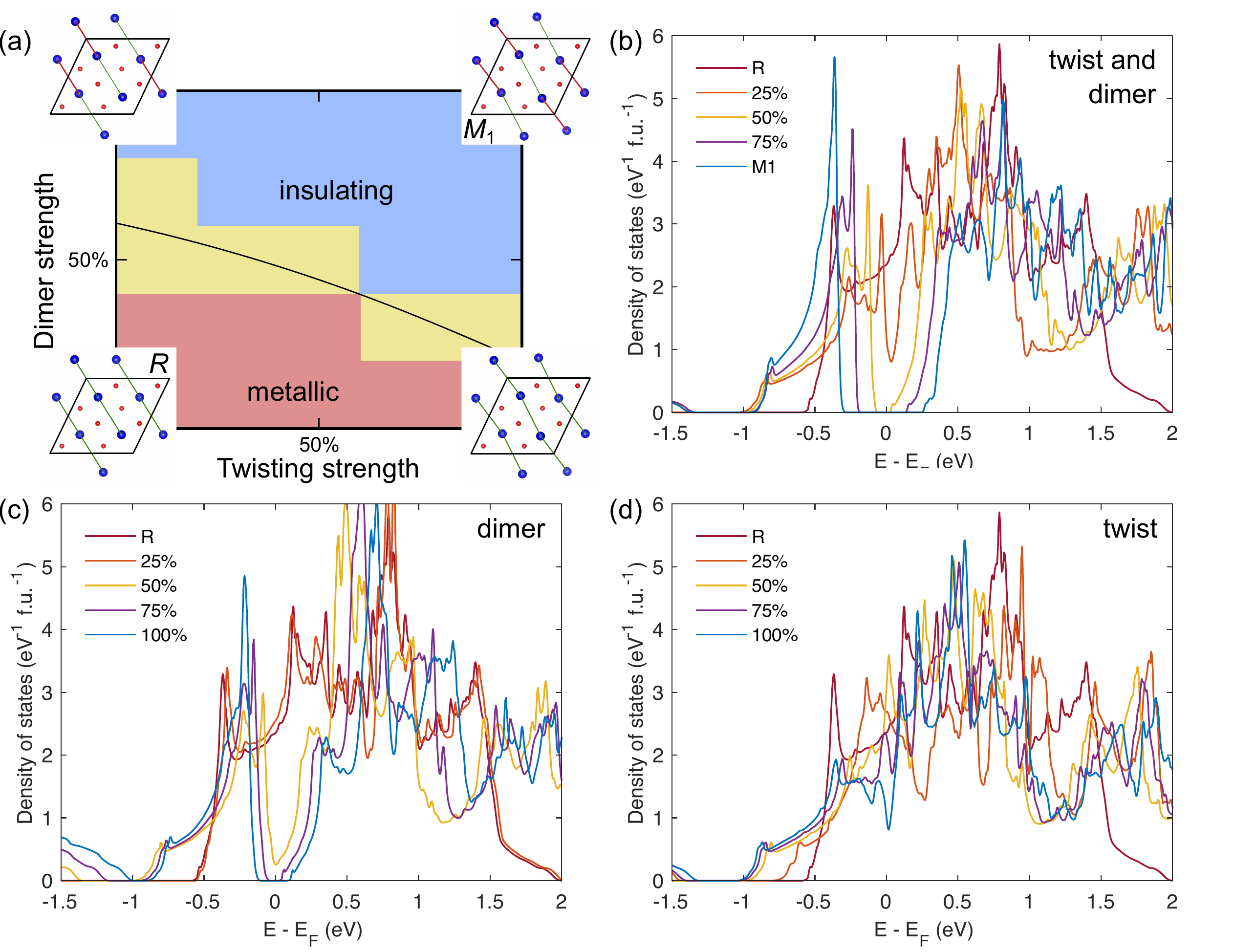}
\end{center}
\vspace*{-0.2in}
\caption{\label{f:fig4} 
Total density of states of VO$_2$ from GGA+U calculations as a function of
independent evolution in the bond length and monoclinic distortion (twisting of
the dimer and VO$_6$ octahedra).
(a)~Solutions of calculations, with
illustrations depicting the nature of the two distortions.
(b)~Variations in both V-V bond length and
monoclinic distortion between {\em R} and $M_1$ phases.
(c)~Variations in the dimer strength (V-V bond length) only,
without monoclinic distortion.
(d)~Variations in the twisting of the VO$_6$ octahedra (monoclinic distortion)
only for constant V-V bond length.
In each figure, 100\% indicates the $M_1$ value.}
\end{figure}

By examining the correlation between the XAS-PEEM image at the $d_{\|}$ energy
and the LEEM image, we are able to assess how closely related the spatial
variations in crystal (LEEM) and dimerization (XAS-PEEM) structures
are;\cite{laverock2014,mcleod2017} auto- and cross-correlation coefficients at
different temperatures are shown in Fig.~\ref{f:fig3}.
Perpendicular to the stripe axis, the images [panels (a-c)] show strong
oscillatory behavior typical of stripe order,\cite{laverock2014} from which we
estimate the stripe widths as $235 \pm 12$ (51~$^{\circ}$C), $207 \pm 18$
(58~$^{\circ}$C), and $170 \pm 33$~nm (69~$^{\circ}$C),
measured from the mean first
zero-crossing (with standard error). Note that the damping of these oscillations
along the perpendicular axis is similar in each correlation coefficient,
suggesting equivalent long-range order in both LEEM and XAS-PEEM contrast
mechanisms. At 51~$^{\circ}$C [Fig.~\ref{f:fig3}(d)],
there is excellent agreement between structures
in the LEEM and XAS-PEEM auto- and cross-correlations, indicating that the two
contrast mechanisms are strongly coupled. However, at 58~$^{\circ}$C
[Fig.~\ref{f:fig3}(e)], these begin to diverge and, crucially, are qualitatively
different from one another. For example, the second satellites
are found at 1.28~{\textmu}m in the LEEM [Fig.~\ref{f:fig3}(a)]
compared with 870~nm in the XAS-PEEM [Fig.~\ref{f:fig3}(c)]
auto-correlations. These are separated from the first satellites (which
represent the mean stripe separation, 580~nm from LEEM and 520~nm from XAS-PEEM)
by a negative trough, which has a specific
meaning here.\cite{supplementary}
Similar behavior is observed at 69~$^{\circ}$C [Fig.~\ref{f:fig3}(f)].
This disparity in the structure of the correlation above 58~$^{\circ}$C is not
expected if the mechanism for contrast in the two measurement techniques is
sensitive to the same underlying structure, i.e.~in our case, if the crystal
symmetry and V-V dimerization are directly and
causally linked. Nevertheless, it is clear from Fig.~\ref{f:fig3}, and in
Figs.~\ref{f:fig1} and \ref{f:masks} above, that sufficiently similar patterns
are observed in both images that the two contrasts are related. We therefore
conclude that the symmetry and V-V dimerization are different,
albeit related, responses to the structural and electronic instabilities that
develop in the system in this intermediate part of the phase diagram.

Given the above results, we can establish a detailed picture of the progression
of the MIT in strained VO$_2$(110). Initially, VO$_2$ is insulating and resides
in the $M_1$ structure, as is well known. With raising temperature, the strict
$M_1$ structure is quickly lost, as the dimerization of the V ions along the
rutile $c_{\rm R}$-axis is weakened, though the system remains electronically
insulating. By 51~$^{\circ}$C, the system separates
into more strongly and weakly
dimerized V-V bonds, yielding XAS-PEEM stripe contrast at the $d_{\|}$
energy.
Initially, the spatial variation in the magnitudes of the dimer
strength and the monoclinic distortion are strongly coupled to
one another. On further increasing the temperature, the V-V bonds are weakened
further, resulting in a decrease in intensity of the XAS-PEEM near the
$d_{\|}$ photon energy. By 58~$^{\circ}$C, the system transitions to a
metallic state, but this transition does not follow the underlying stripe
structure. Rather, we suggest that once a critical mean V-V bond length is
reached, the metallic phase is stabilized and propagates rapidly through the
sample. Although there exist short-range correlations between
the monoclinic distortion and V-V dimerization, these are not commensurate with
one another over long length-scales.
Above this temperature, however, monoclinic-like stripes remain over a wide
temperature range, corresponding to weak dimers in a distorted rutile structure.
In terms of the mechanism of the MIT, our key observation is that the metallic
phase is not correlated with the stripe structure, but rather advances in a
narrow temperature range independent of the variations in crystal structure,
which is easily reconciled in the presence of strong electron correlations.

It is pertinent to ask what role, if any, other lower
symmetry phases of VO$_2$ might play in our
interpretation.\cite{park2013,gu2010} It
is well known that the monoclinic $M_2$ phase, for example, consists of one half
of the V-V ions arranged in a twisted (undimerized) 1D chain and the other half
arranged in a dimerized (untwisted) chain (compared with fully dimerized and
twisted V-V bonds in $M_1$ VO$_2$).\cite{pouget1974} However,
the $M_2$ phase is an
insulator with a sizeable gap, and experimental XAS spectra bear little
resemblance to those of our intermediate
phase.\cite{quackenbush2016} The triclinic {\em T} phase,
although insulating with a band gap comparable to the $M_1$ and $M_2$
phases,\cite{zylbersztejn1975,huffman2017} is transitional between these two structures, whereby
the equally paired and twisted dimers in $M_1$ gradually and continuously
differentiate into the structures in $M_2$.\cite{pouget1974} The {\em T} phase
illustrates the inherent structural instability of the V-V dimers, and that a
continuous evolution in their strength and orientation towards a gradual
breaking of these bonds is already well established in VO$_2$, which has
recently been suggested to play a role in the $M_1 \rightarrow R$ transition of
bulk VO$_2$.\cite{yao2010}

In order to test our interpretations, we have performed electronic structure
calculations of VO$_2$ using the {\sc Elk} code\cite{dewhurst2016} within the
GGA+U formalism ($U=2.8$~eV),\cite{yao2010} explicitly separating the two
structural components of the MIT into independent evolutions in the V-V bond
length and the twisting of the dimer and VO$_6$ octahedra within the unit
cell.\cite{supplementary}
All solutions using the rutile V-V bond length (i.e.\ un-dimerized) are
metallic,
Fig.~\ref{f:fig4}(a,d).
On the other hand, varying the bond lengths between those
of the rutile and $M_1$ phases leads to an insulating solution for dimer
strengths
of 50\% and greater towards the $M_1$ structure [Fig.~\ref{f:fig4}(c)]. The
significance of this result is that for ``weak dimerization'', characterized by
dimers of $\leq 25$\% of the $M_1$ bond strengths, GGA+U predicts that VO$_2$ is
metallic, irrespective of the twisting of the dimer and VO$_6$ octahedra
[Figs.~\ref{f:fig4}(b,d)]. In the presence of biaxial strain, these intermediate
structures may be metastable during the progress of the MIT, in qualitative
agreement with our XAS-PEEM results.

Without strong electron correlations, the band gap in insulating $M_1$ VO$_2$ is
anticipated to be small.\cite{zylbersztejn1975,eyert2002et} Partly,
this may be explained by the reduced screening available in the $M_1$ phase due
to the emptying of the $\pi^*$ ($e_g^{\pi}$) orbital and the narrowing of the
occupied
$d_{\|}$ ($a_{1g}$) due to a reduction in mixing.\cite{zylbersztejn1975} It is
therefore noteworthy that contrast spectra (i.e.~the difference between ``dark
LEEM'' and ``bright LEEM'' spectra) based on our XAS-PEEM results indicate a
broader width ($\approx 1.4$~eV) at $\geq 58$~$^{\circ}$C (i.e.~in the metallic
phases). In the insulating phases ($\leq 51$~$^{\circ}$C), the contrast spectra
have a width of $\approx 1.0$~eV, comparable to the difference spectrum of $(M_1
- R)$. Together with the details of the stripes discussed above, these results
represent strong evidence that electron correlations are central to the MIT
mechanism in VO$_2$.

In summary, the
$M_1$ VO$_2$ is quickly destroyed on heating,
and is replaced by an insulating monoclinic phase that gradually
separates into strongly- and weakly-paired dimers that are coupled to the
underlying symmetry of the crystal structure.
The electronic transition occurs over a narrow temperature
range within this structural evolution, giving way to a monoclinic-like metallic
phase that still contains bimodal dimer pairing and symmetry.
The continuous weakening of the structural distortion
eventually establishes the rutile
phase. These details are inconsistent with a structural MIT
model, depending instead on the effects of strong electron
correlation.

\section*{Acknowledgments}
The Boston University program is supported in part by the Department of Energy
under Grant No.\ DE-FG02-98ER45680. S.K., J.W.L.\ and S.A.W.\ gratefully
acknowledge financial support from the Army Research Office through MURI grant
No.\ W911-NF-09-1-0398. J.L.\ acknowledges the EPSRC Centre for Doctoral
Training in Condensed Matter Physics (CDT-CMP), Grant No.\ EP/L015544/1.
Calculations were performed using the computational facilities of the Advanced
Computing Research Centre, University of Bristol (http://www.bris.ac.uk/acrc/).
The underlying research materials can be accessed at the following DOI:
\href{http://dx.doi.org/10.5523/xxx}{10.5523/xxx}.


\begin{thebibliography}{99}

\bibitem{laverock2014}
J.\ Laverock, S.\ Kittiwatanakul, A.\ A.\ Zakharov, Y.\ R.\ Niu, B.\ Chen,
S.\ A.\ Wolf, J.\ W.\ Lu, and K.\ E.\ Smith,
Direct Observation of Decoupled Structural and Electronic Transitions and an
Ambient Pressure Monocliniclike Metallic Phase of VO$_2$,
\href{http://dx.doi.org/10.1103/PhysRevLett.113.216402}
{Phys.\ Rev.\ Lett.\ {\bfseries 113}, 216402 (2014)}.

\bibitem{arcangeletti2007et}
E.\ Arcangeletti, L.\ Baldassarre, D.\ Di Castro, S.\ Lupi, L.\ Malavasi,
C.\ Marini, A.\ Perucchi, and P.\ Postorino,
Evidence of a Pressure-Induced Metallization Process in Monoclinic VO$_2$,
\href{http://dx.doi.org/10.1103/PhysRevLett.98.196406}
{Phys.\ Rev.\ Lett.\ {\bfseries 98}, 196406 (2007)};
W.-P.\ Hsieh, M.\ Trigo, D.\ A.\ Reis, G.\ A.\ Artioli, L.\ Malavasi, and
W.\ L.\ Mao,
Evidence for photo-induced monoclinic metallic VO$_2$ under high pressure,
\href{http://dx.doi.org/10.1063/1.4862197}
{Appl.\ Phys.\ Lett.\ {\bfseries 104}, 021917 (2014)}.

\bibitem{kim2008et}
B.-J.\ Kim, Y.\ W.\ Lee, S.\ Choi, J.-W.\ Lim, S.\ J.\ Yun, H.-T.\ Kim,
T.-J.\ Shin, and H.-S.\ Yun,
Micrometer x-ray diffraction study of VO$_2$ films: Separation between
metal-insulator transition and structural phase transition,
\href{http://dx.doi.org/10.1103/PhysRevB.77.235401}
{Phys.\ Rev.\ B {\bfseries 77}, 235401 (2008)};
Z.\ Tao, T.-R.\ T.\ Han, S.\ D.\ Mahanti, P.\ M.\ Duxbury, F.\ Yuan,
C.-Y.\ Ruan, K.\ Wang, and J.\ Wu,
Decoupling of Structural and Electronic Phase Transitions in VO$_2$,
\href{http://dx.doi.org/10.1103/PhysRevLett.109.166406}
{Phys.\ Rev.\ Lett.\ {\bfseries 109}, 166406 (2012)};
T.\ L.\ Cocker, L.\ V.\ Titova, S.\ Fourmaux, G.\ Holloway, H.-C.\ Bandulet,
D.\ Brassard, J.-C.\ Kieffer, M.\ A.\ El Khakani, and F.\ A.\ Hegmann,
Phase diagram of the ultrafast photoinduced insulator-metal transition in
vanadium dioxide,
\href{http://dx.doi.org/10.1103/PhysRevB.85.155120}
{Phys.\ Rev.\ B {\bfseries 85}, 155120 (2012)}.

\bibitem{morin1959et}
F.\ J.\ Morin,
Oxides which show a metal-to-insulator transition at the N\'{e}el temperature,
\href{http://dx.doi.org/10.1103/PhysRevLett.3.34}
{Phys.\ Rev.\ Lett.\ {\bfseries 3}, 34 (1959)};
N.\ F.\ Mott, {\em Metal-Insulator Transitions}
(Taylor \& Francis Ltd, London, 1974).

\bibitem{goodenough1973}
J.\ B.\ Goodenough and H.\ Y.-P.\ Hong,
Structures and a two-band model for the system V$_{1-x}$Cr$_x$O$_2$,
\href{http://dx.doi.org/10.1103/PhysRevB.8.1323}
{Phys.\ Rev.\ B {\bfseries 8}, 1323 (1973)}.

\bibitem{jeong2013}
J.\ Jeong, N.\ Aetukuri, T.\ Graf, T.\ D.\ Schladt, M.\ G.\ Samant, and
S.\ S.\ P.\ Parkin,
Suppression of Metal-Insulator Transition in VO$_2$ by Electric Field-Induced
Oxygen Vacancy Formation,
\href{http://dx.doi.org/10.1126/science.1230512}
{Science {\bfseries 339}, 1402 (2013)}.

\bibitem{cavalleri2001}
A.\ Cavalleri, C.\ T\"{o}th, C.\ W.\ Siders, and J.\ A.\ Squier,
Femtosecond Structural dynamics in VO$_2$ during an ultrafast solid-solid phase
transition,
\href{http://dx.doi.org/10.1103/PhysRevLett.87.237401}
{Phys.\ Rev.\ Lett.\ {\bfseries 87}, 237401 (2001)}.

\bibitem{park2013}
J.\ H.\ Park, J.\ M.\ Coy, T.\ S.\ Kasirga, C.\ Huang, Z.\ Fei, S.\ Hunter, and
D.\ H.\ Cobden,
Measurement of a solid-state triple point at the metal-insulator transition in
VO$_2$,
\href{http://dx.doi.org/10.1038/nature12425}
{Nature {\bfseries 500}, 431 (2013)}.

\bibitem{nakano2012et}
M.\ Nakano, K.\ Shibuya, D.\ Okuyama, T.\ Hatano, S.\ Ono, M.\ Kawasaki,
Y.\ Iwasa, and Y.\ Tokura,
Collective bulk carrier delocalization driven by electrostatic surface charge
accumulation,
\href{http://dx.doi.org/10.1038/nature11296}
{Nature {\bfseries 487}, 459 (2012)};
K.\ Appavoo, B.\ Wang, N.\ F.\ Brady, M.\ Seo, J.\ Nag, R.\ P.\ Prasankumar,
D.\ J.\ Hilton, S.\ T.\ Pantelides, and R.\ F.\ Haglund, Jr.,
Ultrafast Phase Transition via Catastrophic Phonon Collapse Driven by
Plasmonic Hot-Electron Injection,
\href{http://dx.doi.org/10.1021/nl4044828}
{Nano Lett.\ {\bfseries 14}, 1127 (2014)};
K.\ Liu, C.\ Cheng, Z.\ Cheng, K.\ Wang, R.\ Ramesh, and J.\ Wu,
Giant-Amplitude, High-Work Density Microactuators with Phase Transition
Activated Nanolayer Bimorphs,
\href{http://dx.doi.org/10.1021/nl303405g}
{Nano Lett.\ {\bfseries 12}, 6302 (2012)}.

\bibitem{muraoka2002}
Y.\ Muraoka and Z.\ Hiroi,
Metal-insulator transition of VO$_2$ thin films grown on TiO$_2$(001) and (110)
substrates,
\href{http://dx.doi.org/10.1063/1.1446215}
{Appl.\ Phys.\ Lett.\ {\bfseries 80}, 583 (2002)}.

\bibitem{cao2009}
J.\ Cao, E.\ Ertekin, V.\ Srinivasan, W.\ Fan, S.\ Huang, H.\ Zheng,
J.\ W.\ L.\ Yim, D.\ R.\ Khanal, D.\ F.\ Ogletree, J.\ C.\ Grossman, and J.\ Wu,
Strain engineering and one-dimensional organization of metal-insulator domains
in single-crystal vanadium dioxide beams,
\href{http://dx.doi.org/10.1038/nnano.2009.266}
{Nat.\ Nanotechnol.\ {\bfseries 4}, 732 (2009)}.

\bibitem{haverkort2005}
M.\ W.\ Haverkort, Z.\ Hu, A.\ Tanaka, W.\ Reichelt, S.\ V.\ Streltsov,
M.\ A.\ Korotin, V.\ I.\ Anisimov, H.\ H.\ Hsieh, H.-J.\ Lin, C.\ T.\ Chen,
D.\ I.\ Khomskii, and L.\ H.\ Tjeng,
Orbital-assisted metal-insulator transition in VO$_2$,
\href{http://dx.doi.org/10.1103/PhysRevLett.95.196404}
{Phys.\ Rev.\ Lett.\ {\bfseries 95}, 196404 (2005)}.

\bibitem{zylbersztejn1975}
A.\ Zylbersztejn and N.\ F.\ Mott,
Metal-insulator transition in vanadium dioxide,
\href{http://dx.doi.org/10.1103/PhysRevB.11.4383}
{Phys.\ Rev.\ B {\bfseries 11}, 4383 (1975)}.

\bibitem{pouget1974}
J.\ P.\ Pouget, H.\ Launois, T.\ M.\ Rice, P.\ Dernier, A.\ Gossard,
G.\ Villeneuve, and P.\ Hagenmuller,
Dimerization of a linear Heisenberg chain in the insulating phases of
V$_{1-x}$Cr$_x$O$_2$,
\href{http://dx.doi.org/10.1103/PhysRevB.10.1801}
{Phys.\ Rev.\ B {\bfseries 10}, 1801 (1974)}.

\bibitem{pouget1975et}
J.\ P.\ Pouget, H.\ Launois, J.\ P.\ D'Haenens, P.\ Merenda, and T.\ M.\ Rice,
Electron localization induced by uniaxial stress in pure VO$_2$,
\href{http://dx.doi.org/10.1103/PhysRevLett.35.873}
{Phys.\ Rev.\ Lett.\ {\bfseries 35}, 873 (1975)};
G.\ Villeneuve, A.\ Bordet, A.\ Casalot, J.\ P.\ Pouget, H.\ Launois, and
P.\ Lederer,
Contribution to the study of the metal-insulator transition in the
V$_{1-x}$Nb$_x$O$_2$ system: I - crystallographic and transport properties,
\href{http://dx.doi.org/10.1016/S0022-3697(72)80494-3}
{J.\ Phys.\ Chem.\ Solids {\bfseries 33}, 1953 (1972)}.

\bibitem{gu2010}
Y.\ Gu, J.\ Cao, J.\ Wu, and L.-Q.\ Chen,
Thermodynamics of strained vanadium dioxide single crystals,
\href{http://dx.doi.org/10.1063/1.3499349}
{J.\ Appl.\ Phys.\ {\bfseries 108}, 083517 (2010)}.

\bibitem{budai2014}
J.\ D.\ Budai, J.\ Hong, M.\ E.\ Manley, E.\ D.\ Specht, C.\ W.\ Li,
J.\ Z.\ Tischler, D.\ L.\ Abernathy, A.\ H.\ Said, B.\ M.\ Leu,
L.\ A.\ Boatner, R.\ J.\ McQueeney, and O.\ Delaire,
Metallization of vanadium dioxide driven by large phonon entropy,
\href{http://dx.doi.org/10.1038/nature13865}
{Nature {\bfseries 515}, 535 (2014)}.

\bibitem{huffman2017}
T.\ J.\ Huffman, C.\ Hendriks, E.\ J.\ Walter, J.\ Yoon, H.\ Ju, R.\ Smith,
G.\ L.\ Carr, H.\ Krakauer, and M.\ M.\ Qazilbash,
Insulating phases of vanadium dioxide are Mott-Hubbard insulators,
\href{http://dx.doi.org/10.1103/PhysRevB.95.075125}
{Phys.\ Rev.\ B {\bfseries 95}, 075125 (2017)}.

\bibitem{kumar2014}
S.\ Kumar, J.\ P.\ Strachan, M.\ D.\ Pickett, A.\ Bratkovsky, Y.\ Nishi, and
R.\ S.\ Williams,
Sequential Electronic and Structural Transitions in VO$_2$ Observed Using
X-ray Absorption Spectromicroscopy,
\href{http://dx.doi.org/10.1002/adma.201402404}
{Adv.\ Mater.\ {\bfseries 26}, 7505 (2014)}.

\bibitem{laverock2012}
J.\ Laverock, A.\ R.\ H.\ Preston, J.\ Newby, D., K.\ E.\ Smith, S.\ Sallis,
L.\ F.\ J.\ Piper, S.\ Kittiwatanakul, J.\ W.\ Lu, S.\ A.\ Wolf,
M.\ Leandersson, and T.\ Balasubramanian,
Photoemission evidence for crossover from Peierls-like to Mott-like transition
in highly strained VO$_2$,
\href{http://dx.doi.org/10.1103/PhysRevB.86.195124}
{Phys.\ Rev.\ B {\bfseries 86}, 195124 (2012)}.

\bibitem{kittiwatanakul2014et}
S.\ Kittiwatanakul, S.\ A.\ Wolf, and J.\ W.\ Lu,
Large epitaxial bi-axial strain induces a Mott-like phase transition in VO$_2$,
\href{http://dx.doi.org/10.1063/1.4893326}
{Appl.\ Phys.\ Lett.\ {\bfseries 105}, 073112 (2014)};
M.\ Yang, Y.\ Yang, B.\ Hong, L.\ Wang, K.\ Hu, Y.\ Dong, H.\ Xu, H.\ Huang,
J.\ Zhao, H.\ Chen, L.\ Song, H.\ Ju, J.\ Zhu, J.\ Bao, X.\ Li, Y.\ Gu,
T.\ Yang, X.\ Gao, Z.\ Luo, and C.\ Gao,
Suppression of Structural Phase Transition in VO$_2$ by Epitaxial Strain in
Vicinity of Metal-insulator Transition,
\href{http://dx.doi.org/10.1038/srep23119}
{Sci.\ Rep.\ {\bfseries 6}, 23119 (2016)}.

\bibitem{yao2010}
T.\ Yao, X.\ Zhang, Z.\ Sun, S.\ Liu, Y.\ Huang, Y.\ Xie, C.\ Wu, X.\ Yuan,
W.\ Zhang, Z.\ Wu, G.\ Pan, F.\ Hu, L.\ Wu, Q.\ Liu, and S.\ Wei,
Understanding the Nature of the Kinetic Process in a VO$_2$ Metal-Insulator
Transition,
\href{http://dx.doi.org/10.1103/PhysRevLett.105.226405}
{Phys.\ Rev.\ Lett.\ {\bfseries 105}, 226405 (2010)}.

\bibitem{abbate1991et}
M.\ Abbate, F.\ M.\ F.\ de Groot, J.\ C.\ Fuggle, Y.\ J.\ Ma, C.\ T.\ Chen,
F.\ Sette, A.\ Fujimori, Y.\ Ueda, and K.\ Kosuge,
Soft-x-ray-absorption studies of the electronic-structure changes through the
VO$_2$ phase transition,
\href{http://dx.doi.org/10.1103/PhysRevB.43.7263}
{Phys.\ Rev.\ B {\bfseries 43}, 7263 (1991)};
T.\ C.\ Koethe, Z.\ Hu, M.\ W.\ Haverkort, C.\ Sch\"{u}{\ss}ler-Langeheine,
F.\ Venturini, N.\ B.\ Brookes, O.\ Tjernberg, W.\ Reichelt, H.\ H.\ Hsieh,
H.-J.\ Lin, C.\ T.\ Chen, and L.\ H.\ Tjeng,
Transfer of spectral weight and symmetry across the metal-insulator
transition in VO$_2$,
\href{http://dx.doi.org/10.1103/PhysRevLett.97.116402}
{Phys.\ Rev.\ Lett.\ {\bfseries 97}, 116402 (2006)}.

\bibitem{laverock2012b}
J.\ Laverock, L.\ F.\ J.\ Piper, A.\ R.\ H.\ Preston, B.\ Chen, J.\ McNulty,
K.\ E.\ Smith, S.\ Kittiwatanakul, J.\ W.\ Lu, S.\ A.\ Wolf, P.-A.\ Glans, and
J.-H.\ Guo,
Strain dependence of bonding and hybridization across the metal-insulator
transition of VO$_2$,
\href{http://dx.doi.org/10.1103/PhysRevB.85.081104}
{Phys.\ Rev.\ B {\bfseries 85}, 081104(R) (2012)}.

\bibitem{supplementary}
See Supplementary Information at [url].

\bibitem{bauer1994}
E.\ Bauer,
Low energy electron microscopy,
\href{http://dx.doi.org/10.1088/0034-4885/57/9/002}
{Rep.\ Prog.\ Phys.\ {\bfseries 57}, 895 (1994)}.

\bibitem{bauer2001}
E.\ Bauer,
Photoelectron spectromicroscopy: present and future,
\href{http://dx.doi.org/10.1016/S0368-2048(00)00261-9}
{J.\ Electron Spectrosc.\ Rel.\ Phenom.\ {\bfseries 114}, 975 (2001)}.

\bibitem{west2008}
K.\ G.\ West, J.\ Lu, J.\ Yu, D.\ Kirkwood, W.\ Chen, Y.\ Pei, J.\ Claassen, and
S.\ A.\ Wolf,
Growth and characterization of vanadium dioxide thin films prepared by
reactive-biased target ion beam deposition,
\href{http://dx.doi.org/10.1116/1.2819268}
{J.\ Vac.\ Sci.\ Technol.\ A {\bfseries 26}, 133 (2008)}.

\bibitem{kittiwatanakul2013}
S.\ Kittiwatanakul, J.\ Laverock, D.\ Newby, Jr., K.\ E.\ Smith,
S.\ A.\ Wolf, and J.\ Lu,
Transport behavior and electronic structure of phase pure VO$_2$ thin films
grown on $c$-plane sapphire under different O$_2$ partial pressure,
\href{http://dx.doi.org/10.1063/1.4817174}
{J.\ Appl.\ Phys.\ {\bfseries 114}, 053703 (2013)}.

\bibitem{laverock2015}
J.\ Laverock, S.\ Kittiwatanakul, A.\ A.\ Zakharov, Y.\ R.\ Niu, B.\ Chen,
J.\ Kuyyalil, S.\ A.\ Wolf, J.\ W.\ Lu, and K.\ E.\ Smith,
Simultaneous Spectroscopic, Diffraction and Microscopic Study of the
Metal-Insulator Transition of VO$_2$,
\href{http://dx.doi.org/10.1557/opl.2015.480}
{Mater.\ Res.\ Soc.\ Symp.\ Proc.\ {\bfseries 1730}, 480 (2015)}.

\bibitem{contrast}
Note that the contrast in Fig.~\ref{f:fig1}(b,e,h)
is inverted with respect to the LEEM images, indicating
that higher absorption at $d_{\|}$ is associated with the monoclinic-like
structural phase observed in LEEM, in agreement with
XAS.\cite{abbate1991et,laverock2012b}

\bibitem{mcleod2017}
A.\ S.\ McLeod, E.\ van Heumen, J.\ G.\ Ramirez, S.\ Wang, T.\ Saerbeck,
S.\ Guenon, M.\ Goldflam, L.\ Anderegg, P.\ Kelly, A.\ Mueller, M.\ K.\ Liu,
I.\ K.\ Schuller, and D.\ N.  Basov,
Nanotextured phase coexistence in the correlated insulator V$_2$O$_3$,
\href{http://dx.doi.org/10.1038/nphys3882}
{Nat.\ Phys.\ {\bfseries 13}, 80 (2017)}.

\bibitem{quackenbush2016}
N.\ F.\ Quackenbush, H.\ Paik, M.\ J.\ Wahila, S.\ Sallis, M.\ E.\ Holtz,
X.\ Huang, A.\ Ganose, B.\ J.\ Morgan, D.\ O.\ Scanlon, Y.\ Gu, F.\ Xue,
L.-Q.\ Chen, G.\ E.\ Sterbinsky, C.\ Schlueter, T.-L.\ Lee, J.\ C.\ Woicik,
J.-H.\ Guo, J.\ D.\ Brock, D.\ A.\ Muller, D.\ A.\ Arena, D.\ G.\ Schlom, and
L.\ F.\ J.\ Piper,
Stability of the $M_2$ phase of vanadium dioxide induced by coherent epitaxial
strain,
\href{http://dx.doi.org/10.1103/PhysRevB.94.085105}
{Phys.\ Rev.\ B {\bfseries 94}, 085105 (2016)}.

\bibitem{eyert2002et}
V.\ Eyert,
The metal-insulator transitions of VO$_2$: A band theoretical approach,
\href{http://dx.doi.org/10.1002/1521-3889(200210)11:9<650::AID-ANDP650>3.0.CO;2-K}
{Ann.\ Phys.\ (Leipzig) {\bfseries 11}, 650 (2002)};
S.\ Biermann, A.\ Poteryaev, A.\ I.\ Lichtenstein, and A.\ Georges,
Dynamical singlets and correlation-assisted Peierls transition in VO$_2$,
\href{http://dx.doi.org/10.1103/PhysRevLett.94.026404}
{Phys.\ Rev.\ Lett.\ {\bfseries 94}, 026404 (2005)};
B.\ Lazarovits, K.\ Kim, K.\ Haule, and G.\ Kotliar,
Effects of strain on the electronic structure of VO$_2$,
\href{http://dx.doi.org/10.1103/PhysRevB.81.115117}
{Phys.\ Rev.\ B {\bfseries 81}, 115117 (2010)};
V.\ Eyert,
VO$_2$: A novel view from band theory,
\href{http://dx.doi.org/10.1103/PhysRevLett.107.016401}
{Phys.\ Rev.\ Lett.\ {\bfseries 107}, 016401 (2011)};
W.\ H.\ Brito, M.\ C.\ O.\ Aguiar, K.\ Haule, and G.\ Kotliar,
Metal-Insulator Transition in VO$_2$: A DFT+DMFT Perspective,
\href{http://dx.doi.org/10.1103/PhysRevLett.117.056402}
{Phys.\ Rev.\ Lett.\ {\bfseries 117}, 056402 (2016)};
O.\ N\'{a}jera, M.\ Civelli, V.\ Dobrosavljevi\'{c}, and M.\ J.\ Rozenberg,
Resolving the VO$_2$ controversy: Mott mechanism dominates the
insulator-to-metal transition,
\href{http://dx.doi.org/10.1103/PhysRevB.95.035113}
{Phys.\ Rev.\ B {\bfseries 95}, 035113 (2017)}.

\bibitem{dewhurst2016}
J.\ K.\ Dewhurst, S.\ Sharma, L.\ Nordst\"{o}m, F.\ Cricchio, F.\ Bultmark,
and E.\ K.\ U.\ Gross,
The Elk FP-LAPW code,
\href{http://elk.sourceforge.net}
{\tt{http://elk.sourceforge.net}} (2016).

\end{thebibliography}
\end{document}